\documentclass{article}
\usepackage{graphicx,amssymb}

\begin{document}
\title{Quasi-stationary states in low-dimensional Hamiltonian systems}
\author{Fulvio Baldovin,
Edgardo Brigatti
and Constantino Tsallis\thanks{E-mail addresses: baldovin@cbpf.br, edgardo@cbpf.br, 
tsallis@cbpf.br}\\
\it{Centro Brasileiro de Pesquisas F\'{\i}sicas}\\
\it{Rua Xavier Sigaud 150, 
22290-180 Rio de Janeiro-RJ, Brazil.}
}
\maketitle

\begin{abstract}
We address a simple connection between   
results of Hamiltonian nonlinear dynamical theory
and thermostatistics.  
Using a properly defined dynamical temperature in low-dimensional symplectic maps,
we display and characterize long-standing 
quasi-stationary states that eventually cross over to a 
Boltzmann-Gibbs-like regime. As time evolves, the geometrical properties 
(e.g., fractal dimension) of the phase space
change sensibly,
and the duration of the anomalous regime diverges with decreasing chaoticity.
The scenario that emerges is consistent with the nonextensive statistical mechanics one.
\end{abstract}


The methods of usual, Boltzmann-Gibbs (BG), statistical
mechanics apply to impressively 
large classes of macroscopic systems.
However, the situation is more delicate for complex systems.
Indeed, turbulent
fluids \cite{Beck_01}, high-energy collision processes
\cite{Bediaga_01}, classical \cite{Baldovin_01} and quantum 
chaos \cite{chaos}, stellar self-gravitating systems
\cite{Taruya_01}, granular systems \cite{Gheorghiu_01},
economics \cite{Borland_01}, motion of 
micro-organisms \cite{Upadhyaya_01}, and others,
frequently exhibit anomalous behaviors where alternative 
approaches are needed.
In particular, in 
many-body long-range-interacting Hamiltonian systems,
it has been recently observed the emergence of long-standing 
(in the thermodynamical limit {\it infinite-lasting})
{\it quasi-stationary} 
({\it metastable}) {\it states} (QSS)
characterized by non-Gaussian velocity distributions, 
before the Boltzmann-Gibbs (BG) equilibrium is attained 
\cite{Latora_01,Nobre_01}. 
This is a major concern, as, for these Hamiltonian systems,
the foundation of the BG equilibrium thermodynamics is 
questioned.
Using standard results, in this letter we
address a simple connection between chaos theory and 
thermostatistics, and we focus on a paradigmatic 
dynamical mechanism that produces QSS very similar to those 
detected in \cite{Latora_01,Nobre_01}. 
These QSS are displayed and
characterized by means of low-dimensional symplectic maps.

The foundation of the Boltzmann-Gibbs (BG) equilibrium
thermodynamics lies on a sufficiently complete and 
uniform occupation of the system phase space (the {\it finite} Lebesgue measure  
$\Gamma$-space), taking into account symmetry, energy and
similar restrictions.  
The BG equilibrium descends in fact from the {\it
equal-a-priori-probability postulate}, that characterizes
the microcanonical Gibbsian ensemble (see, e.g.,
\cite{Huang_01}).  
According to this postulate, each
equally-sized accessible region of the phase space (under
the macroscopic conditions of the system) equally likely
contains the microscopic state of the system.  As Einstein
pointed out in his criticism of the Boltzmann principle $S=k
\ln W$ \cite{Einstein_01}, this postulate should not be
taken {\it a priori}, but rather justified 
{\it a posteriori} by the underlying dynamics.
Indeed, if dynamics is sufficiently chaotic, large
portion of the phase space are rapidly occupied by the
trajectory of the system and the postulate is a very accurate
representation of the dynamical behavior, as testifies more
than a century of successes of the BG formalism.  
But there are also many situations where the system 
displays an intricate dynamical behavior, as it happens 
for example at the border between regular and chaotic regimes.

At this border,  for a large class of Hamiltonian systems, 
a mechanism based on the KAM theory operates, which we briefly review now.
A continuous Hamiltonian system with $n$ degrees of freedom may be
written in the form \cite{Zaslavsky_01}:
\begin{equation}
H=H_0(I_1,...,I_n)+\epsilon\;V(I_1,\theta_1,...,I_n,\theta_n),
\end{equation}  
where $H_0$ is integrable ($I_1,...,I_n$ are its integrals of motion), 
$\epsilon<<1$, and $V$ is a nonlinear perturbation.
Under certain hypothesis (see, e.g., \cite{Zaslavsky_01}), 
for $\epsilon=0$ the trajectories lie on invariant $n$-dimensional tori.
A special subset of these tori are called {\it resonance tori}.
Specifically, 
if we introduce the (non-degenerate) frequencies of the unperturbed motion:
$\omega_j\equiv\frac{\partial H_0}{\partial I_j}\;(j=1,...,n)$,
we have that the condition
$\sum_{j=1}^n m_j\omega_j=0$
(where $m_j$ are integer numbers)
defines the resonance tori.
Each resonance torus involves the formation of a separatrix loop.
The action of the perturbation, for small enough 
$\epsilon\neq 0$, deforms normal tori into KAM-tori,
and, in correspondence with the resonance tori,  
destroys the separatrices replacing them with {\it stochastic layers}.
Resonance tori, in the space spanned by $\omega_1,..,\omega_n$, lie
in the intersection between the hyperplane defined by the resonance condition
and the hypersurface of energy $E=H_0(\omega_1,...,\omega_n)$.
In the case $n>2$, resonance tori must, for topological reasons, intersect 
between them. Consequently, while for $n\leq 2$ the stochastic layers are distinct
for $\epsilon$ sufficiently small, for $n>2$ they merge into a single 
connected stochastic web that is dense in the phase space 
for all $\epsilon\neq 0$, and there is room for Arnold diffusion processes.
We remark that, for $n=2$, KAM-tori constitute {\it total barriers} for diffusive processes
in the phase space; nevertheless, inside the stochastic sea, it is possible
to find Cantor sets, named {\it cantori}, 
that constitute {\it partial barriers} 
for diffusion (see \cite{MacKay_01} for details).

A convenient way of studying Hamiltonian systems is by using {\it symplectic maps}.
A $(2n-2)$-dimensional symplectic map is
obtained from {\it conservative} Hamiltonian systems 
with $n$ degrees of freedom
by taking a Poincar\'e section over the hypersurface of constant energy.
Interestingly enough, a
$2n$-dimensional symplectic map is also the result of
a Poincar\'e section on the phase space of an {\it open}
system of $n$ degrees of freedom with a Hamiltonian that
depends {\it periodically} on time. 
We remark that in {\it both} cases the map has a
symplectic structure; this assures (hyper)volume
conservation in the phase space.
The advantage of maps lies on the reduced dimension of the phase 
space and on the use of a discrete time.
In this letter we specifically address some 
symplectic (hence conservative) maps in order to discuss
how equilibrium and quasi-equilibrium can be attained in 
phase space.

Let us start through the analysis of a prototypical $2$-dimensional
symplectic map ($n\leq 2$), 
the {\it standard} (or  {\it kicked rotor}) {\it map} 
\begin{eqnarray}
\theta(t+1)&=&p(t)+\frac{a}{2\pi}\sin[2\pi \theta(t)]+\theta(t)~~~\rm{(mod\;1)},
\label{standard}\\
p(t+1)&=&p(t)+\frac{a}{2\pi}\sin[2\pi \theta(t)]~~~~~~~~~\rm{(mod\;1)}\nonumber
\end{eqnarray}
$(a\in{\mathbb R},\;\;t=0,1,...)$.
$2\pi p$ may be regarded as an angular momentum variable.
Notice that, consistently with our scope, 
we have used the symmetry 
properties of the map and defined, as usually, the angular momentum $\rm{mod\;1}$.
The standard map is integrable for $a=0$, while chaoticity rapidly increases
with $|a|$.

\begin{figure}
\begin{center}
\includegraphics[width=1.\textwidth,angle=0]{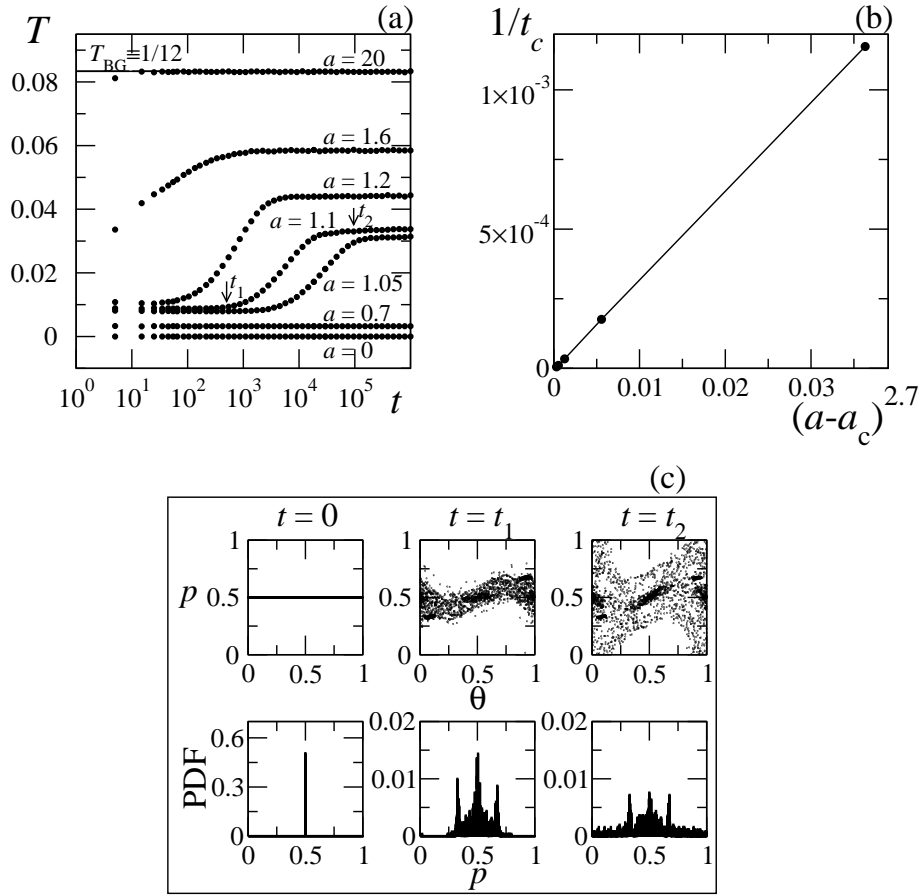}
\end{center}
\caption{\small 
(a) Time evolution of the dynamical temperature $T$ 
of a standard map, for typical values of $a$. We start with `water bag' 
initial conditions ($M=2500$ points in  $0\leq\theta\leq 1$, $p=0.5\pm 5\;10^{-4}$).
In order to eliminate cyclical fluctuations, the dots represent
average of $10$ iteration steps; moreover, each curve is
the average of $50$ realizations. 
(b) Inverse crossover time $t_c$ (inflection point
between the QSS and the BG regimes)
vs. $1/(a-a_c)^{2.7}$. No inflection points subsist if $t$ is linearly represented. 
(c) Time evolution of the the ensemble in (a)
for $a=1.1$ (first row) and 
PDF of its angular momentum (second row). 
$t=0$: `water bag' initial conditions; $t=t_1=500$: the ensemble 
is mostly restricted by cantori; 
$t=t_2=10^5$: the ensemble is confined inside KAM-tori.
} 
\label{fig_std_01}
\end{figure}

In \cite{Latora_01,Nobre_01}, the emergence of the dynamical
QSS appeared to be dependent 
on the initial conditions.
Specifically, it was shown that for some classical long-range-interacting
$N$-rotor Hamiltonian models,
a basin of attraction of initial data exist for which the system 
dynamically evolves into a QSS whose duration diverges as $N\to\infty$.
Typical examples of this basin of attraction are
out-of-equilibrium initial conditions called `water bag' initial
conditions, characterized by a uniform initial 
distribution of the angular momenta around zero (see
\cite{Latora_01,Nobre_01} for details).
In the case of the standard map, we first observe that the points $(0,1/2)$ and 
$(1/2,1/2)$ are a $2$-cycle for all $a$ so that we can use
them as referential for studying the properties of the phase
space with respect to variation of the
parameter $a$. 
With some analogy with \cite{Latora_01,Nobre_01}, our out-of-equilibrium 
`water bag' initial conditions are defined  
by considering at $t=0$ a 
{\it statistical ensemble} of $M$ copies of the standard map 
with arbitrary $\theta$ and $p$ randomly distributed in a small region around
$p=1/2$. 
In standard statistical mechanics, when dealing with systems with diagonal 
kinetic matrix and zero average 
momentum, the temperature is proportional to the average square
momentum per particle.
As we analyze situations with nonzero `bulk' motion,
the analogous concept, which we shall refer to as 
(dimensionless) `dynamical temperature',
can be defined  
as the {\it variance} of the angular momentum: 
$T\equiv \langle (p-\langle p \rangle)^2 \rangle      
=\langle p^2 \rangle -\langle p \rangle^2$, 
where $\langle \rangle$ means  ensemble average. 
The temperature associated with the uniform ensemble 
(that we will call BG temperature because of its similarity
with the equal-a-priori-probability postulate) is given by
$T_{\rm{BG}}\equiv\int_0^1 dp\;p^2-\left(\int_0^1dp\;p\right)^2=1/12$.
It should be noticed that in the present conservative model, the `temperature' $T$ is
necessarily bounded since 
$p$ itself is bounded, in contrast with a true thermodynamical
temperature, which is of course unbounded. 
For large values of $|a|$ (i.e., strong chaoticity) in map (\ref{standard}), the temperature of the `water bag' initial ensemble  
rapidly relaxes to $T_{\rm{BG}}$ (let us stress that the subindex $BG$ stands for the fact that it corresponds to {\it uniform} occupation of the accessible phase space of the map, to be distinguished from the phase space of the physical kicked rotor from which the map was originally deduced). 
Our aim is to study
what happens in the transition to regularity obtained
reducing the value of $|a|$ towards $a=0$.
In Fig. \ref{fig_std_01}(a) we see that 
the first effect of the reduction of $|a|$ 
is that $\lim_{ t \to \infty} T(t) < T_{\rm{BG}}$. This is easily understood as follows.
As chaoticity reduces, total barriers (KAM-tori) appear in the phase space.
As a consequence, the points of the ensemble are prevented to reach all the
regions of the phase space and
the projection of the ensemble on the $p$ axis produces
a probability distribution function (PDF) with a variance smaller than 
the one of the uniform distribution. 
For values of $a$ of order $a\sim 1$,
a QSS  emerges before the relaxation to the final temperature. 
In fact, inside the stochastic sea partial barriers 
(cantori) begins to appear, and 
the initial `water bag' first rapidly diffuse inside an area delimited by
cantori and then slowly crosses over to the final relaxation temperature.
Fig. \ref{fig_std_01}(c) illustrates this behavior.
To obtain a quantitative description of the relaxation time we have 
plotted the iteration time in a logarithmic scale. We define the crossover
time $t_c$ as the inflection point of the curve and we observe that it
diverges, as 
$a$ tends to $a_c=0.971635406...$ from above, 
like $t_c\sim1/(a-a_c)^{2.7}$ (see Fig. \ref{fig_std_01}(b)). 
Just below this critical value in fact, the strongest cantori close
\cite{MacKay_01}, and the relaxation to a higher temperature
is prevented.
Reducing $a$ to smaller values causes the formation of more and more total barriers,
so that the `dynamical temperature' tends to zero for $|a|\to 0$.
>From the mechanism we have displayed, it is clear that it is  
possible to obtain these types of QSS  
even with other sets of initial conditions. 
Typically, it is sufficient to have the initial data 
localized inside the first partial barriers. 
In other words there is an entire basin of attraction of 
out-of-equilibrium initial conditions that leads to the 
formation of a certain kind of QSS.

As we pointed out previously, the topology of the phase space 
changes dramatically for $n>2$. To address this case, we move next to a $4$-dimensional
symplectic map composed by two coupled standard maps:
\begin{eqnarray}
\theta_1(t+1)&=&p_1(t+1)+\theta_1(t)+b\;p_2(t+1),\nonumber\\
p_1(t+1)&=&p_1(t)+\frac{a_1}{2\pi}\sin[2\pi \theta_1(t)],
\label{standard_standard}\\
\theta_2(t+1)&=&p_2(t+1)+\theta_2(t)+b\;p_1(t+1),\nonumber\\
p_2(t+1)&=&p_2(t)+\frac{a_2}{2\pi}\sin[2\pi \theta_2(t)],\nonumber
\end{eqnarray}
where $a_1,a_2,b\in{\mathbb R},\;t=0,1,...$,  and all variables are defined 
$\rm{mod\;1}$. If the coupling constant $b$ vanishes the two standard maps decouple;
if $b=2$ the
points $(0,1/2,0,1/2)$ and $(1/2,1/2,1/2,1/2)$ are a $2$-cycle for all
$(a_1, a_2)$, hence we preserve in phase space the same referential  that we had for a single 
standard map. For a generic value of $b$, all relevant present results remain qualitatively the same.           
Also, we set $a_1=a_2\equiv\tilde a$ so that the system is invariant under
permutation $1\leftrightarrow 2$. Since we have two rotors now, the  
`dynamical temperature' is naturally given by
$T\equiv \frac{1}{2}
\left(<p_1^2>+<p_2^2>-<p_1>^2-<p_2>^2\right)$,
hence the BG temperature remains $T_{\rm{BG}}\equiv 1/12$.

\begin{figure}
\begin{center}
\includegraphics[width=1.\textwidth,angle=0]{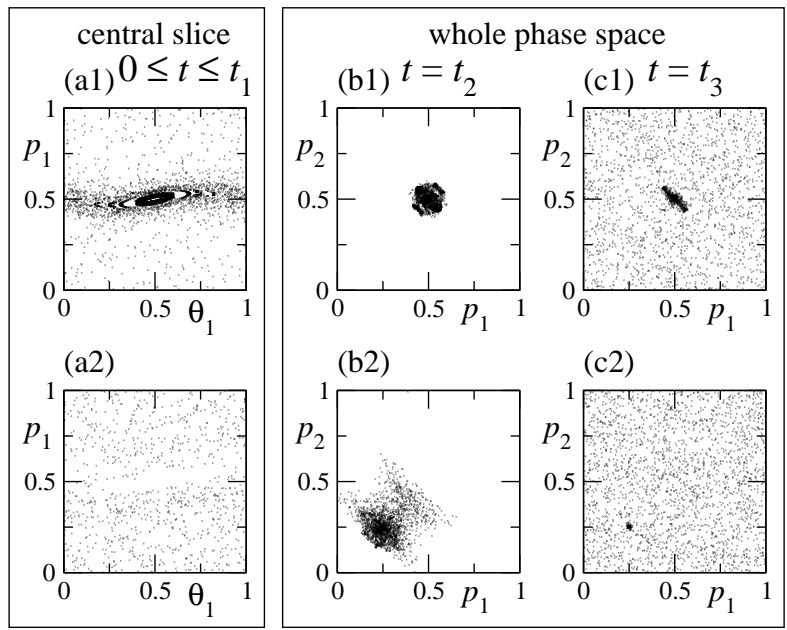}
\end{center}
\caption{\small 
Phase space analysis of the evolution of `water bag' ensembles
for two coupled standard maps for $(\tilde a,b)=(0.4,2)$.
First row: `Water bag' 
initial conditions $0\leq\theta_1,\theta_2\leq 1$, 
$p_1,p_2=0.5\pm 5\;10^{-3}$.
Second row: `Water bag' 
initial conditions $0\leq\theta_1,\theta_2\leq 1$, 
$p_1,p_2=0.25\pm 5\;10^{-3}$.
(a) Projection on the $(\theta_1,p_1)$-plane of the central slice of the
phase space ($\theta_2,p_2=0.5\pm 10^{-2}$), 
for the orbit $0\leq t\leq t_1=10^4$. 
(c),(c) Projection on the $(p_1,p_2)$-plane of whole phase space 
for the iterate at time $t_2=15$ and $t_3=2\;10^4$.
} 
\label{fig_std_std_01}
\end{figure}

As before we consider `water bag' initial conditions, i.e., an
ensemble of $M$ points with arbitrary $(\theta_1$, $\theta_2)$, and angular
momenta randomly distributed inside a small region around $p_1= p_2=1/2$. The result is qualitatively similar to the one displayed
in Fig. \ref{fig_std_01}(a) for $a>a_c$.
Large values of $\tilde a$ correspond to $T_{{\rm BG}}$ and reducing 
$\tilde a$ we observe the formation of a QSS that, after some time, relaxes to
a temperature $T<T_{{\rm BG}}$.
The first major difference with the 2-dimensional case is that, 
because of 
the Arnold diffusion processes, the relaxation to a higher temperature
occurs (waiting enough time) for {\it all} $\tilde a\neq 0$ (i.e., $\tilde a_c=0$, defining 
$\tilde a_c$ as the value where $t_c$ diverges).
Moreover, the reason why the ensemble does not relax to the BG 
temperature for small values of $\tilde a$ is here quite different from that for
the $2$-dimensional case. Indeed, 
with this choice of initial data, the initial `water bag'
intersects at least a macroscopic island, as clearly appreciated in 
Fig. \ref{fig_std_std_01} (a1); the points that are set inside the island do not diffuse to the outside.
As a result, the projection of the ensemble on the plane $p_1,p_2$ conserve  
a denser central part for all times (Fig. \ref{fig_std_std_01} (b1), (c1)).

\begin{figure}
\begin{center}
\includegraphics[width=1.\textwidth,angle=0]{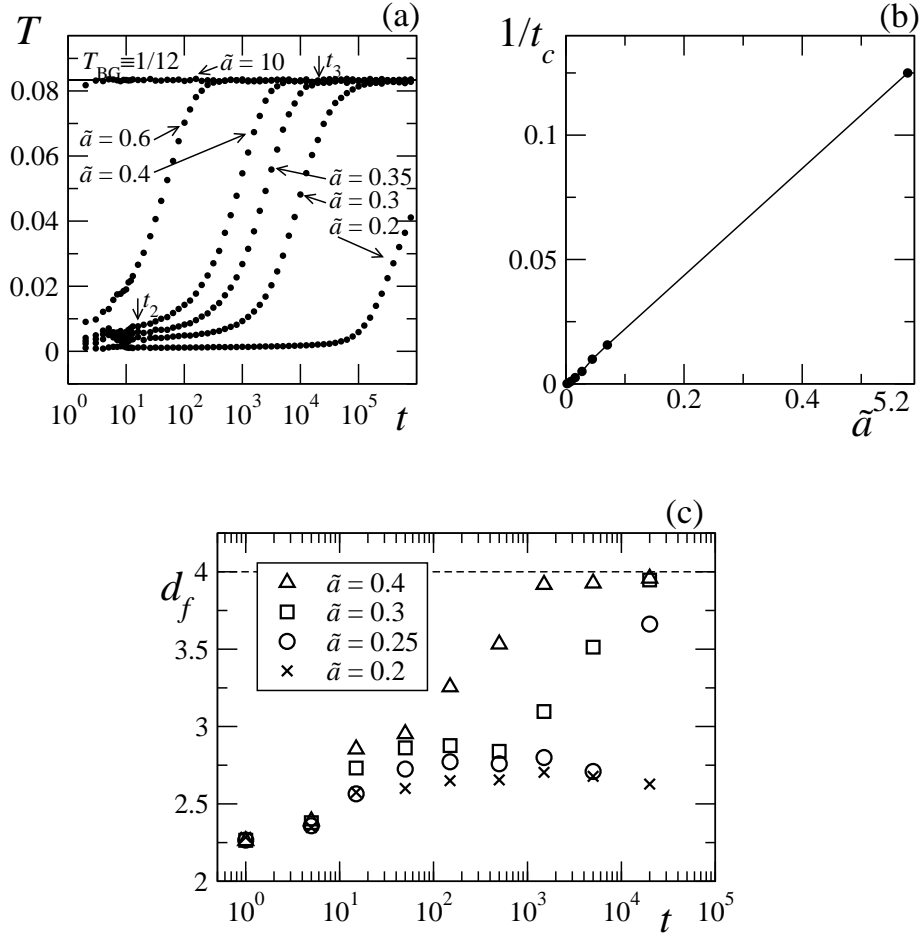}
\end{center}
\caption{\small
(a) Time evolution of the dynamical temperature $T$ 
of two coupled standard maps, for $b=2$ and typical values of $\tilde a$. 
We start with water bag 
initial conditions ($M=1296$ points 
with 
$0\leq\theta_1,\theta_2\leq 1$, and 
$p_1,p_2=0.25\pm 5\;10^{-3}$);
moreover, an average was taken over $35$ realizations.
See Fig. \ref{fig_std_std_01} for $t_2$ and $t_3$.
(b) Inverse crossover time $t_c$
vs. $1/\tilde a^{5.2}$. 
(c) Time evolution of the fractal dimension 
of a single initial ensemble in the same 
setup of (a).
} 
\label{fig_std_std_02}
\end{figure}

If we instead shift the initial `water bag', say towards the lower part of the
phase space, we can set the points outside this island 
(Fig. \ref{fig_std_std_01} (a2)-(c2)).
In this case we obtain a crucial qualitatively new phenomenon, namely
the formation, for small values of $\tilde a$, of a QSS 
that eventually relaxes to the BG temperature
(see Fig. \ref{fig_std_std_02}(a)). Notice also that the crossover
time $t_c$ diverges, as $\tilde a\to 0$, faster than in the
$2$-dimensional case: $t_c\sim1/\tilde a^{5.2}$ 
(see Fig. \ref{fig_std_std_02}(b)). 
We remark that the relaxation to the BG temperature
occurs here {\it even} if the presence of islands in the phase space
violate the equal a priori
postulate. 
In other words, it is possible to obtain a   
{\it weak} violation of the postulate that {\it does} preserve
a uniform distribution once the ensemble is projected 
over the plane $p_1,p_2$, in the same sense that 
a sponge projects a uniform shadow on a wall.
These QSS can in fact be geometrically characterized by the {\it fractal dimension} $d_f$.
Overcoming some numerical difficulties involved
in a fractal analysis in $4$ dimensions, we illustrate what happens 
in Fig. \ref{fig_std_std_02}(c), constructed using a box-counting algorithm \cite{kruger_01} 
(in fact, the phase space exhibits strong inhomogeneities which suggest a multifractal structure). 
During the QSS the ensemble 
is, for small $\tilde a$, associated with a nontrivial
fractal dimension $d_f \simeq 2.7$, while, once it crosses over
to the BG-like regime, it distributes itself occupying the full
dimensionality of the phase space, thus attaining $d_f=4$.

\begin{figure}
\begin{center}
\includegraphics[width=1.\textwidth,angle=0]{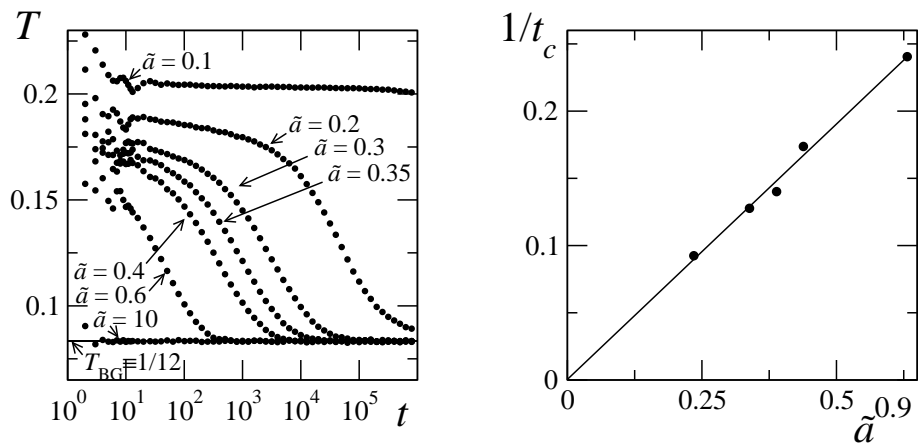}
\end{center}
\caption{\small 
Same as Fig. \ref{fig_std_std_02}(a),(b)
but with `double water bag' initial conditions:
\mbox{$0\leq\theta_1,\theta_2\leq 1$}; 
$p_1,p_2$ randomly distributed 
inside one of the two regions
$p_1,p_2=0+10^{-2}$, $p_1,p_2=1-10^{-2}$.
} 
\label{fig_std_std_03}
\end{figure}

These simple models   
allows also for the discussion of different types of QSS. For instance, 
at $t=0$ we can set up `double water bag' initial conditions
considering an ensemble of $M$ copies of two coupled standard maps
with arbitrary $\theta_1,\theta_2$ and angular momenta 
randomly distributed inside {\it two} small regions: 
$p_1,p_2=0+\delta$ and $p_1,p_2=1-\delta$ ($0<\delta <<1$). 
In this case, the initial temperature $T(0)$ is higher than $T_{\rm{BG}}$, 
because the PDFs projected on the $p_1,p_2$ axes are double peaked. 
Relaxation to $T_{\rm{BG}}$ occurs then {\it from above},
as can be seen in Fig. \ref{fig_std_std_03}.
This is precisely in what the  
phenomenon observed in \cite{Nobre_01} differs from
the one in \cite{Latora_01}.

It is relevant to notice that, as a consequence of Arnold diffusion,
our results for the $4$-dimensional map capture the qualitative behavior
of higher-dimension symplectic maps. In fact, the two-plateaux structure
has been confirmed by numerical analyses where hundreds of standard maps
are all-with-all coupled as in Eq. (\ref{standard_standard}) 
\cite{Majtey_01}.
  
Through the numerical analysis of low-dimensional symplectic
maps, we have displayed how complex paradigmatic structures associated with  
conservative nonlinear dynamics can generate anomalous thermodynamical behavior.
Particularly, we have exhibited and studied the emergence, while approaching integrability (i.e., when chaoticity decreases), 
of QSS suggestively similar to those observed in long-range $N$-body systems 
\cite{Latora_01,Nobre_01} ($\tilde a$ playing a role analogous to $1/N$).
A central result is that, in contrast with what happens for the BG equilibrium,
these QSS correspond to a nontrivial fractal dimension. 
This situation reminds the phase space structure of logistic-like maps at
the edge of chaos (also characterized by a nontrivial fractal dimension), 
where exact analytical connections with the nonextensive
statistical mechanics \cite{Tsallis_01} have been established \cite{Baldovin_01}. 

\section*{Acknowledgments} 
We thank C. Anteneodo, A. Kruger, A.P.Majtey, A. Rapisarda, A. Robledo and J. de Souza for
useful remarks, as well as CAPES, PRONEX, CNPq and 
FAPERJ (Brazilian agencies) for partial support.

\newpage

\newpage

\newpage
\vspace{3.0cm}

\end{document}